\documentstyle[12pt]{article}
\font\titlefont=cmbx10 scaled \magstep4

\begin{document}
\input{epsf}

\begin{flushright}
\vspace*{-2cm}
gr-qc/9901074 \\ 
January 26, 1999 \\
\vspace*{1cm}
\end{flushright}

\begin{center}
{\titlefont THE QUANTUM INTEREST CONJECTURE} \\
\vskip .7in
L.H. Ford\footnote{email: ford@cosmos2.phy.tufts.edu}
 and 
Thomas A. Roman\footnote{Permanent address: Department of Physics and Earth
Sciences, Central Connecticut State University, New Britain, CT 06050 \\
email: roman@ccsu.edu} \\
\vskip .2in
Institute of Cosmology\\
Department of Physics and Astronomy\\
Tufts University\\
Medford, Massachusetts 02155\\
\end{center}

\vspace*{1cm}
\begin{abstract}
Although quantum field theory allows local negative energy densities 
and fluxes, it also places severe restrictions upon the magnitude and extent 
of the negative energy. The restrictions take the form of quantum inequalities. 
These inequalities imply that a pulse of negative energy must not only be 
followed by a compensating pulse of positive energy, but that the temporal 
separation between the pulses is inversely proportional to their amplitude. 
In an earlier paper we conjectured that there is a further constraint upon a 
negative and positive energy delta-function pulse pair. This conjecture 
({\it the quantum interest conjecture}) states that a positive energy pulse 
must overcompensate the negative energy pulse by an amount which is 
a monotonically increasing function of the pulse separation. In the present 
paper we prove the conjecture for massless quantized scalar fields in two 
and four-dimensional flat spacetime, and show that it is implied by the 
quantum inequalities. 
\end{abstract}
\newpage

\baselineskip=14pt

\section{Introduction}
\label{sec:intro}
Known forms of classical matter obey the 
weak energy condition (WEC): \break
$T_{\mu\nu} u^{\mu} u^{\nu} \geq 0$, 
where $T_{\mu\nu}$ is the stress-energy 
tensor of matter, and $u^{\mu}$ is an arbitrary 
timelike vector \cite{HE}. By continuity, 
the condition also holds for all null vectors. 
Physically, the WEC implies that the energy 
density seen by any observer must be 
non-negative. However, the renormalized 
stress-energy tensors of quantum fields 
can violate the WEC, as well as all other 
known pointwise energy conditions \cite{EGJ,Kuo}. 
States of quantum fields which involve negative 
energy densities have even been produced in the 
laboratory, two examples being the Casimir effect 
\cite{C,L,MR,HL} and squeezed states of light 
\cite{WK}, although the energy densities have 
not been directly measured. If 
there were no constraints on negative energy 
densities, then it would be possible to use 
them to produce macroscopic effects. Such 
effects might include warp drives \cite{A,K}, 
traversable wormholes \cite{MT}, time machines 
\cite{MTY,HCP,E}, and violations of the second law of 
thermodynamics \cite{F78,D82}. The extent to which 
the laws of physics place restrictions on negative energy 
density has received much attention during the last few years. 

Recent progress has been made on the topic of 
``quantum inequalities.'' These are inequalities 
which restrict the magnitude 
and duration of negative energy densities and 
fluxes \cite{F78,F91,FR90,FR92,FR95,FR97}. 
Physically, the inequalities imply that the energy density 
seen by an observer cannot be arbitrarily 
negative for an arbitrarily long period of time 
\cite{COMMENT:CAS-EFFECT}. 
The mathematical form of the bound consists of 
the renormalized expectation value of the energy 
density or the flux, evaluated in an arbitrary 
quantum state, and folded into a sampling 
function. The latter is a peaked function of time 
with a characteristic width, $t_0$, called the 
sampling time. For a quantized massless, 
minimally coupled scalar field in
four-dimensional Minkowski spacetime, the original 
bound on the energy density has the 
form \cite{FR95,FR97}: 
\begin{equation}
\hat \rho \equiv {{t_0} \over \pi}\, 
\int_{-\infty}^{\infty}\,
{{T_{\mu\nu} u^{\mu} u^{\nu}\, dt}
\over {t^2+{t_0}^2}} \geq
-{3\over {32 {\pi}^2 {t_0}^4}}\,,  \label{eq:4DENQI}
\end{equation}
for all sampling times $t_0$, where 
$t$ is the proper time of a geodesic observer. 
(Throughout this paper we use units in which 
$\hbar=c=1$.) Equation~(\ref{eq:4DENQI}) was 
derived using the Lorentzian sampling function 
given by 
\begin{equation}
g(t) = \frac{t_0}{ \pi \,(t^2 + {t_0}^2)} \,.
\label{eq:lorentz}
\end{equation}

Quantum inequalities 
have now been derived in curved as 
well as flat spacetimes, in both two and four 
dimensions \cite{PF971,Song,PFGQI,FT}. They 
have also been used to obtain severe constraints 
on traversable wormhole geometries \cite{FRWH} 
and warp drives \cite{PFWD,ER}. Quantum 
inequalities have been derived for massless 
and massive scalar fields 
\cite{FR95,FR97,PF971,PFGQI,FT}, and for the 
electromagnetic field \cite{FR97}. 

Flanagan \cite{FLAN} generalized the original 
two-dimensional flat spacetime inequalities to 
those with arbitrary sampling functions and was 
also able to find the optimal bound. His result is 
\begin{equation}
{\hat \rho}_{min} = - \frac{1}{24 \,\pi} \, 
\int_{-\infty}^{\infty} \, 
\frac{{g'(t)}^2}{g(t)} \, dt  \,,
\label{eq:flanqi}
\end{equation}
where $g(t)$ is an arbitrary sampling function. 
If the Lorentzian sampling function is substituted 
into Eq.~(\ref{eq:flanqi}), the resulting bound 
is a factor of six smaller than the original 
two-dimensional bound found in Refs. \cite{FR95,FR97}. 

Fewster and Eveson (FE) \cite{FE} have discovered 
a much simpler method for deriving the 
quantum inequalities than was given previously. 
They prove quantum inequalities for 
massless and massive scalar fields in two and four-
dimensional flat spacetimes. Their result for the 
massless scalar field in four dimensions is given by 
\begin{equation}
\hat \rho \geq -\frac{1}{16 \,{\pi}^2}\,
\int_{-\infty}^{\infty} \,{\bigl( {g^{1/2}}''(t) \bigr)}^2 \, dt \,,
\label{eq:feqi}
\end{equation}
where again $g(t)$ is an arbitrary sampling function. 
When $g(t)$ is chosen to be the Lorentzian sampling 
function, the resulting bound is $9/64$ of that in 
Eq.~(\ref{eq:4DENQI}). 
 
An extremely useful feature of the results of Flanagan 
and of FE is the freedom to 
employ sampling functions with compact support. 
Some use of this type of sampling function has already 
been made in Ref. \cite{FPR}. This freedom will be 
exploited more fully in the current paper. 

In an earlier paper \cite{FR92}, we found evidence to 
suggest that there may be stronger restrictions on 
negative energy densities than those proved to date. 
There we were concerned with the question of whether 
a $\delta$-function pulse of negative energy injected 
into an extreme charged black hole could produce 
an observable violation of cosmic censorship by making 
the mass temporarily less than the charge. 
We showed that any initial negative $(-)$ energy pulse 
had to be accompanied by a subsequent positive $(+)$ 
energy pulse. (The use of $\delta$-function pulses 
would seem to provide an efficient means of separating 
$(-)$ and $(+)$ energy.) Furthermore, we discovered 
that there existed a quantum inequality-type 
constraint on the magnitude of the $(-)$ energy 
pulse and the time separation between the pulses. 
Similar constraints had also been found for pulses 
in flat spacetime \cite{F91}. This inequality appeared 
to exclude the possibility of an unambiguously 
observable violation of cosmic censorship. 

During the course of that investigation we discovered 
that, at least in certain instances, it was not possible 
to make the pulses {\it exactly} compensate one another. 
It appeared that the $(+)$ pulse had to 
{\it overcompensate} the $(-)$ pulse. Furthermore, it 
appeared that the amount of overcompensation 
increased with increasing time separation of the pulses. 
This behavior suggested to us that a more general 
principle might be at work. We called this the 
{\it ``quantum interest conjecture''}, which states that 
an energy ``loan'', i. e., $(-)$ energy, must always be 
``repaid'', i. e., by $(+)$ energy, with an ``interest'' 
which depends on the magnitude and duration of the 
loan. In the present paper, we prove that the conjecture is 
indeed true, at least for $\delta$-function pulses composed of 
massless scalar fields in Minkowski spacetime. 

In Sec.~\ref{sec:mm}, we present a physically transparent 
example of quantum interest in two dimensions, where the 
$\delta$-function pulses are generated by a moving mirror. It 
is shown in Sec.~\ref{sec:Tmaximum} that there exists a 
general constraint on the maximum pulse separation, which 
is derived using a quantum inequality with compactly-
supported sampling functions. In Sec.~\ref{sec:NessQInt} 
we show that quantum interest is required to exist 
for $\delta$-function pulses in two and four-dimensional 
flat spacetime, even for arbitrarily small pulse separations. 
Our conclusions and some remaining 
open questions are discussed in Sec.~\ref{sec:conclusions}.

\section{A Simple Example: The Moving Mirror}
\label{sec:mm}
\begin{figure}
\begin{center}
\leavevmode\epsfysize=11cm\epsffile{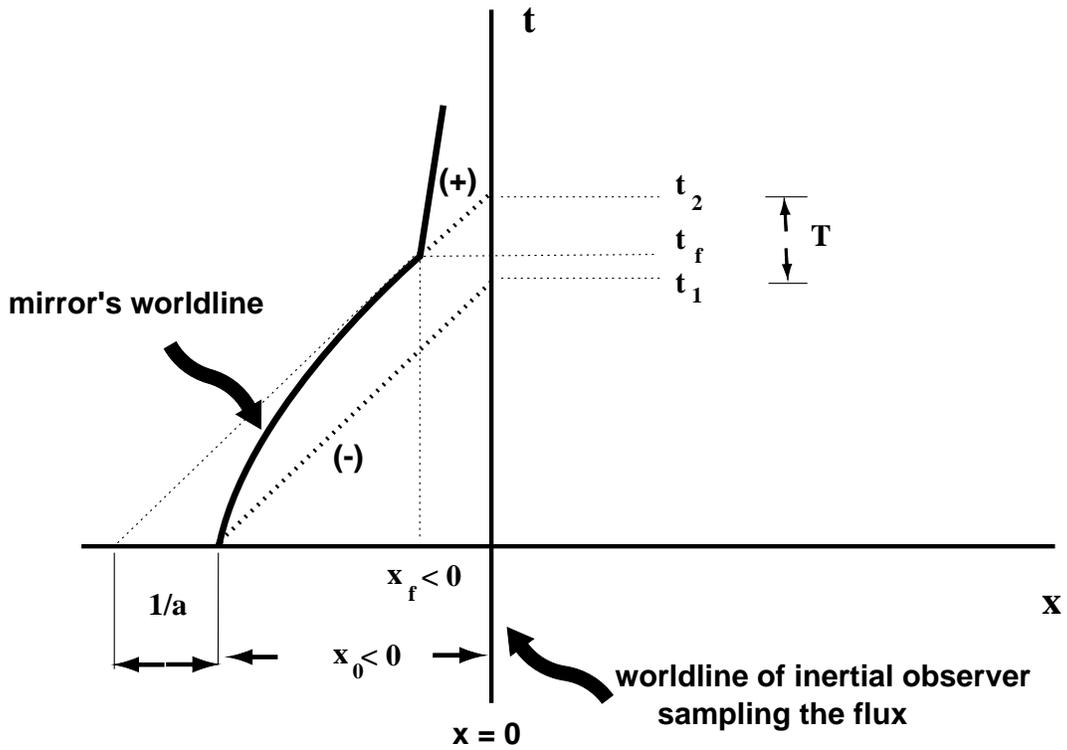}
\end{center}
\caption{
A moving mirror which emits delta function pulses of negative and positive
energy. The mirror is initially at rest at $x =x_0$, and emits a negative
energy pulse as it begins to accelerate with constant 
proper acceleration $a$. At
time $t_f$ and position $x =x_f$, it ceases accelerating and emits a pulse
of positive energy. An inertial observer at $x=0$ receives these pulses
at $t=t_1$ and $t=t_2$, respectively.}
\label{MIRROR}
\end{figure}

As a simple example of quantum interest, we examine 
the case of two $\delta$-function pulses of $(-)$ and $(+)$ energy 
produced by a moving mirror in two-dimensional flat spacetime 
\cite{FD}. Consider an observer at rest at $x=0$, and a mirror which 
accelerates toward the observer from position $x_0<0$. 
(See Fig.~\ref{MIRROR}.) 
The mirror starts from rest and receives an initial kick at 
time $t=0$, which causes it to emit a $\delta$-function pulse 
of $(-)$ energy toward the observer. Subsequently it moves 
with a constant proper acceleration, $a$, until 
$t=t_f$, when its acceleration is abruptly 
halted (to avoid collision with the observer), 
causing the mirror to emit a $\delta$-function 
pulse of $(+)$ energy toward the observer. 
(Unlike a classical point charge, the mirror only 
radiates when its acceleration {\it changes}.) Thereafter 
the mirror moves inertially. The $(-)$ energy pulse crosses 
the observer's worldline at $t=t_1$, and the $(+)$ energy 
pulse crosses at $t=t_2$. Therefore, the pulse separation 
is $T=t_2-t_1$. The mirror's trajectory, during the 
period of constant acceleration, is given by 
\begin{equation}
x(t) = (x_0 - 1/a) + (1/a^2 + t^2)^{1/2} \,.
\label{eq:mtraj}
\end{equation}
The velocity, $V$, of the mirror is given by 
\begin{equation}
V=\frac{dx}{dt}=\frac{at}{(1+a^2 t^2)^{1/2}}\,.
\label{eq:mvel}
\end{equation}

In Ref. \cite{FR90}, it was shown that the 
magnitude of the $(-)$ energy pulse emitted 
by the mirror, $|\Delta E|$, is given by 
\begin{equation}
|\Delta E| = \frac{a}{12 \pi} \,,
\label{eq:DeltaE}
\end{equation} 
and that the magnitude and the pulse separation, $T$, 
are constrained by the inequality 
\begin{equation}
|\Delta E| \, T < \frac{1}{12 \pi} \,.
\label{eq:mDeltaET}
\end{equation}
For an observer to the right of the mirror, the energy 
flux is given by Eq. (2.24) of Ref. \cite{FR90} with 
the sign of $V$ changed. From this expression, 
it can be shown that the magnitude of the positive 
pulse, $\Delta E_p$, is
\begin{equation}
\Delta E_p = \frac{a}{12 \pi} \, 
\frac{\sqrt{1-V_f^2}}{{(1-V_f)}^2} \,,
\label{eq:DeltaEp}
\end{equation}
and therefore 
\begin{equation}
1+ \epsilon = \frac{(\Delta E_p)}{|\Delta E|}=
\frac{\sqrt{1-{V_f}^2}}{(1-V_f)^2} \,,
\label{eq:meps}
\end{equation}
where $\epsilon$ is the fraction by which the 
magnitude of the $(+)$ energy pulse overcompensates 
that of the $(-)$ energy pulse, and $V_f$ is the velocity 
of the mirror when the acceleration ceases. 

We now want to express $\epsilon$ in terms of $|\Delta E|$ 
and $T$. From Fig.~\ref{MIRROR}, it can be seen that 
\begin{equation}
t_2 = t_f + |x_f| = t_f - x_f \,,
\label{eq:t2}
\end{equation}
and
\begin{equation}
t_1=|x_0|=-x_0 \,,
\label{eq:t1}
\end{equation}
where we have used the fact that the final 
position of the mirror when it stops accelerating, 
$x_f$, is negative because the mirror stays to the 
left of the observer. Using $T=t_2-t_1$, and 
Eq.~(\ref{eq:mtraj}), we obtain 
\begin{equation}
t_f = \frac{T \,(aT-2)}{2(aT-1)} \,.
\label{eq:tf}
\end{equation}
Equations~(\ref{eq:mvel}), ~(\ref{eq:meps}), 
~(\ref{eq:tf}), the fact that $aT<1$, and a 
straightforward but slightly tedious calculation 
imply 
\begin{equation}
\epsilon = \frac{aT \, (2 a^2 \, T^2 -5 aT +4)}
                  {2 {(1-aT)}^3} \,.
\label{eq:fmeps}
\end{equation}
If we rewrite this equation using 
Eq.~(\ref{eq:DeltaE}), we get 
\begin{equation}
\epsilon = \frac{24 \,\pi |\Delta E| \,T \, 
(72 \,{\pi}^2 \,{|\Delta E|}^2 \, T^2 -15 \,\pi \,
|\Delta E| \,T +1)}{{(1-12 \, \pi |\Delta E|\, T)}^3} \,,
\label{eq:fmepsE}
\end{equation} 
which is a monotonically increasing function of $|\Delta E| \,T $. 
In the nonrelativistic, i. e., small $|\Delta E| \,T$, limit  
Eq.~(\ref{eq:fmepsE}) becomes  
\begin{equation}
\epsilon \approx  24 \pi \, |\Delta E| \, T +
504 {\pi}^2 \, {(|\Delta E| \, T)}^2 + O [T^3] \dots \,.
\label{eq:nonreleps}
\end{equation}
The leading term on the right-hand side of this expression is identical 
to an earlier result derived in Sec. IIE of 
Ref. \cite{FR92}. There only the nonrelativistic case 
was considered and the more general result, now 
confirmed by our Eq.~(\ref{eq:fmepsE}) above, 
was conjectured.

In this simple example, it is easy to see 
why there is a maximum pulse separation, 
$T<T_{max} = 1/(12 \pi \, |\Delta E|)$, and how 
quantum interest arises. The former arises from 
the constraint that the mirror not collide with the 
observer. The latter is produced by the Doppler shifting 
of the $(+)$ energy pulse due to the nonzero velocity 
of the mirror when the pulse is emitted, 
i. e., at the moment when the mirror's acceleration 
is halted. We speculated in Ref. \cite{FR92} that 
quantum interest might be a more general phenomenon 
which generically occurs in cases of separated 
$\delta$-function pulses of $(-)$ and $(+)$ energy, 
in four as well as in two dimensions, regardless of how 
the pulses are produced. In the following sections of 
this paper, we will prove that this is in fact the case.

\section{A General Constraint on Pulse Separations}
\label{sec:Tmaximum}

In this section, we will show that quantum inequalities 
impose a maximum time separation, $T_{max}$, 
on any pair of $(-)$ and $(+)$ energy 
$\delta$-function pulses, in both two and four-dimensional 
flat spacetime. Let the pulse profile be given by 
\begin{equation}
\rho(t) = B \, [-\delta (t) + (1+\epsilon) \, \delta (t-T)] \,,
\label{eq:pprofile}
\end{equation}
where $B= |\Delta E|$ in two dimensions, and 
$B=|\Delta E|/A$ in four dimensions, where $A$ 
is the (planar) collecting area of the flux \cite{AREA}.

The quantum inequalities have the general form 
\begin{equation}
\hat \rho = \int_{-\infty}^{\infty} \, g(t) \, \rho (t)
\geq -\frac{C}{{t_0}^D} \,,
\label{eq:genrho}
\end{equation}
where $\rho(t)$ is the energy density, 
$g(t)$ is the sampling function, $C$ is a 
constant whose value depends on both the specific 
choice of sampling function and the spacetime 
dimension, $D$. Substituting Eq.~(\ref{eq:pprofile}) into 
Eq.~(\ref{eq:genrho}), we obtain
\begin{equation}
\hat \rho = B \, [-g(0) + (1+\epsilon) \, g(T)] 
\geq -\frac{C}{{t_0}^D} \,.
\label{eq:rhohat}
\end{equation}

Let us use a compactly-supported sampling function 
with width $t_0$, e. g., the sampling function 
vanishes (continuously) for $t \leq -t_0/2$ and for $t \geq t_0/2$. We 
will also assume, for the purposes of this discussion, 
that the sampling function has only one maximum in 
this interval. The left-hand side of 
Eq.~(\ref{eq:rhohat}) will be most negative when 
$T \geq t_0/2$, since $g(T) =0$ in this region. 
The bound can then be written as 
\begin{equation}
B \leq \frac{C}{g(0) \, {t_0}^D} \,.
\label{eq:genBa}
\end{equation}
We may express $t_0$ as a multiple of $T$:
\begin{equation}
t_0 = \kappa T \,,
\end{equation}
where $\kappa \leq 2$, and then rewrite Eq.~(\ref{eq:genBa}) as 
\begin{equation}
B \leq \frac{C}{g(0) \, {\kappa}^D \, T^D} \,.
\label{eq:genBb}
\end{equation}
The best bound is obtained for $\kappa=2$, and is given by 
\begin{equation}
B \leq \frac{C}{2^D \,T^D \, g(0)} \,.
\label{eq:genB}
\end{equation}
For a sampling function with a single maximum, 
$g(0) \propto 1/{t_0}$, so let $g(0) = 
C_0/{t_0}=C_0/2T $, where $C_0$ is a constant whose value 
depends only on the form of the chosen sampling 
function (but not on the spacetime dimension, unlike $C$). 
Equation~(\ref{eq:genB}) now yields the 
constraint 
\begin{equation}
B\,T^{D-1} \leq \frac{C}{2^{D-1} \,C_0} \,.
\label{eq:BTconst}
\end{equation}
We thus have 
the following constraints on the pulse 
separation, $T$:
\begin{equation}
T  \leq \frac{C}{2 \, C_0 \,|\Delta E| } \,,
\label{eq:Tmax2d}
\end{equation}
and 
\begin{equation}
T  \leq \frac{1}{2} \,
{\biggl(\frac{C \,A}{C_0 \,|\Delta E| }
\biggr)}^{1/3} \,,
\label{eq:Tmax4d}
\end{equation}
in two and four dimensions, respectively. Therefore, the 
larger the magnitude of the $(-)$ energy pulse, the smaller 
is the allowed time separation between the pulses. 

An example of a compactly-supported sampling 
function,  which was given in Ref. \cite{FPR}, is 
\begin{equation}
g(t) = \left\{\matrix{0  \,, &  \,\, t < - t_0/2 \cr
(1/ t_0) \, 
[1+ {\rm cos}(2 \pi t / t_0) ] \,,
& \, -t_0/2 \leq t \leq t_0/2 \cr
0 \,, &  \,\, t > t_0/2}\right. \,.
\label{eq:cssf1}
\end{equation}
For this function $C_0=2$, and in two dimensions 
$C=\pi/6$, which gives the following constraint on 
the maximum pulse separation $T_{max}$:
\begin{equation} 
T_{max} =  \frac{\pi}{24 \, |\Delta E|}
\approx \frac{0.131}{|\Delta E|} \,.
\label{eq:cssfT1-2d}
\end{equation}
For the same 
function, $C={\pi}^2/16$ in four dimensions, and 
the analogous constraint is :
\begin{equation} 
T_{max} = {\biggl(\frac{\pi}{16} \biggr)}^{2/3} \,
{\biggl( \frac{A}{|\Delta E|} \biggr)}^{1/3}
\approx 0.338\, {\biggl( \frac{A}{|\Delta E|} \biggr)}^{1/3} \,.
\label{eq:cssfT1-4d}
\end{equation}

Note that 
in two dimensions, the bound on $T_{max}$ 
given by Eq.~(\ref{eq:cssfT1-2d}) is larger than the 
bound ${|\Delta E|}T_{max} = 1/(12 \pi) \approx 
0.027$, for the moving 
mirror case. It could be that there are other 
quantum states in two dimensions for which 
the value of $T_{max}$ lies between the moving 
mirror value and the bound given by 
Eq.~(\ref{eq:cssfT1-2d}). Alternatively, it may 
be that the moving mirror case comes close to 
the maximum allowed value of $T_{max}$ for 
any quantum state in two dimensions. If that 
is true, then perhaps other choices of 
compactly-supported sampling functions would 
lead to stronger bounds on $T_{max}$. 
Of the various sampling functions which 
we have examined, the best bound was 
obtained for Eq.~(\ref{eq:cssf1}). At present, this 
remains an open question .

\section{Necessity of Quantum Interest}
\label{sec:NessQInt}
In this section we will establish the existence 
of quantum interest. The first subsection contains a simple 
argument which proves the necessity of quantum interest in 
four-dimensional flat spacetime, using the Lorentzian sampling 
function. The second subsection extends this argument to 
more general sampling functions in both two and four dimensions. 

\subsection{A Simple Argument in Four Dimensions}
\label{sec:simple}
It will now be shown that quantum interest must 
exist for $\delta$-function pulses in four-dimensional 
flat spacetime. Here we will use the Lorentzian 
sampling function. The quantum inequality is 
\begin{equation}
\hat \rho \geq -\frac{C}{{t_0}^4} \,,
\label{eq:ourQI}
\end{equation} 
where $C= 3/(32 \,\pi^2)$ in the earlier version of 
Ref. \cite{FR95}, and $C=27/(2048 \,\pi^2)$ in the 
improved version of FE \cite{FE}. If we substitute 
Eq.~(\ref{eq:pprofile}), with $B=|\Delta E|/A$, into 
Eq.~(\ref{eq:ourQI}), the resulting inequality 
may be rewritten as 
\begin{equation}
\hat \rho = \frac{|\Delta E|}{\pi \,A} \,
\frac{(\epsilon \,\beta^2- 1)}{\beta T \,
(1+\beta^2)} \geq 
-\frac{C}{{\beta}^4 \, T^4} \,,
\label{eq:rhohat4D}
\end{equation}
where
\begin{equation}
\beta = \frac{t_0}{T} \,.
\label{eq:betadef}
\end{equation}
A non-trivial inequality is obtained only for 
$\epsilon \beta^2 -1 < 0$, i. e., 
$0 \leq \beta < 1/\sqrt{\epsilon}$. 
A slight rearrangement of 
Eq.~(\ref{eq:rhohat4D}) gives 
\begin{equation}
|\Delta E| \, T \leq \biggl(\frac{C \pi A}{T^2} 
\biggr) \, F(\beta) \,,
\label{eq:ETG}
\end{equation}
with 
\begin{equation}
F(\beta) = 
\frac{1+\beta^2}{\beta^3 \,(1-\epsilon \beta^2)} \,.
\label{eq:Gdef}
\end{equation}
Note that if we set $\epsilon = 0$, corresponding 
to {\it exactly} compensating pulses, then 
\begin{equation}
F(\beta) = \frac{1+\beta^2}{\beta^3} \rightarrow 0 \,,
\,\, {\rm as} \,\,\beta \rightarrow \infty \,.
\label{eq:Glimit}
\end{equation}
Since the quantum inequality must hold for any 
value of $\beta$, this would imply that 
$|\Delta E| \,T^3 \rightarrow\nobreak 0$, 
i. e., either $|\Delta E| =0$ or $T=0$. In either case, 
there would be no non-trivial pulses. Therefore, 
$\epsilon >0$, i. e., we {\it must} have quantum interest 
in four-dimensional flat spacetime.

To get the tightest bound in Eq.~(\ref{eq:ETG}), we 
should evaluate the right-hand side at the smallest 
value of the function $F(\beta)$. 
A calculation using the computer algebra program 
MACSYMA shows that the minimum 
of $F(\beta)$ is at 
\begin{equation}
\beta=\beta_m = \sqrt{\frac
{\sqrt{(1+\epsilon) (1+ 25 \epsilon) }  +1 - 5 \epsilon}
{6 \epsilon}} \,.
\label{eq:betamin}
\end{equation}
If we let $y(\epsilon)=F(\beta_m(\epsilon))$, then 
$y(\epsilon)$ is a monotonically increasing function 
(as can be shown, for example, by graphing the function). 
Therefore, Eq.~(\ref{eq:ETG}) can be written as 
\begin{equation}
|\Delta E| \, T \leq \biggl(\frac{C \pi A}{T^2} 
\biggr) \, y(\epsilon) \,,
\label{eq:yeps}
\end{equation}
or, inverting the inequality, as 
\begin{equation}
\epsilon \geq y^{-1} \,\biggl(\frac{|\Delta E| \, T^3}
{C \pi A} \biggr) \,.
\label{eq:eps}
\end{equation}
Since $y(\epsilon)$ is a monotonically increasing function, 
so is $y^{-1}$, hence the minimum allowed value of $\epsilon$ 
increases as $|\Delta E| \, T^3$ increases.

In the $\epsilon \rightarrow 0$ limit, $y(\epsilon) \sim 
3  \sqrt{3 \epsilon}/2$, and so Eq.~(\ref{eq:yeps}) 
can be inverted to yield 
\begin{equation}
\epsilon \geq \frac{4}{27} \, {\biggl(\frac{|\Delta E| \, T^3}
{C \pi A} \biggr)}^2 \,,
\label{eq:epsilonto0}
\end{equation}
i. e., $\epsilon$ grows as $T^6$ for fixed $|\Delta E|$ and $A$.

\subsection{A More General Approach to Quantum Interest}
\label{sec:general}

In this section we will develop a more general formalism for obtaining
lower bounds upon $\epsilon$. First, it is convenient to formulate the quantum 
inequalities in terms of sampling functions which are dimensionless functions 
of a dimensionless variable, $z=t/t_0$. Let
\begin{equation}
G(z) = t_0 g(t) \, .
\end{equation}
 The quantum inequalities still take the form of 
Eq.~(\ref{eq:genrho}), where the constant $C$ is
\begin{equation}
C = \frac{1}{24 \,\pi} \, \int_{-\infty}^{\infty} \, 
\frac{[G'(z)]^2}{G(z)} \, dz  \,,  \label{eq:C2D}
\end{equation}
in two dimensions, and
\begin{equation}
C = \frac{1}{16 \,{\pi}^2}\,
\int_{-\infty}^{\infty} \,{\bigl[ {G^{1/2}}''(z) \bigr]}^2 \, dz \,,
\label{eq:C4D}
\end{equation}
in four dimensions.
The lower bound on $\epsilon$ can be obtained from Eq.~(\ref{eq:rhohat}):
\begin{equation}
\epsilon \geq \frac{1}{g(T)} \, \biggl( g(0) - \frac{C}{B\,{t_0}^D} \biggr) -1 
\,. \label{eq:geneps}
\end{equation}
In terms of $G$, this bound is
\begin{equation}
\epsilon \geq \frac{1}{G(x)} \, \biggl[ C_0 - \frac{C}{B}\,
\biggl(\frac{x}{T}\biggr)^{D-1} \biggl] -1 
\,, \label{eq:geneps2}
\end{equation}
where $C_0 = G(0)$ and 
\begin{equation}
x =T/t_0  \,.
\label{eq:xdef}
\end{equation}
We can express this as an upper bound on $T$ for fixed $\epsilon$ as
\begin{equation}
B\, T^{D-1} \leq H(x,\epsilon) \,,
\end{equation}
where
\begin{equation}
 H(x,\epsilon) = \frac{C\, x^{D-1}}{C_0 -(\epsilon+1) G(x)} \,.
\end{equation}
Note that our bound is nontrivial only if $C_0 -(\epsilon+1) G(x) > 0$.

The basic strategy which we will pursue is the following:  first find the value 
$x = x_m$ which minimizes the function $H(x,\epsilon)$ for fixed $\epsilon$.
Then the best bound on $T$ for a given form of $G$ is
\begin{equation}
B\, T^{D-1} \leq y(\epsilon) \equiv H(x_m,\epsilon) \,. \label{eq:Tbound}
\end{equation}
If $y(\epsilon)$ is a monotonically increasing function, then so is its
inverse $y^{-1}$, and we can write the lower bound on $\epsilon$ as
\begin{equation}
\epsilon \geq y^{-1} (B\, T^{D-1}) \,.
\end{equation}
The extremization with respect to $x$ is equivalent to that with respect to 
$\beta$ performed in the previous section, and amounts to finding the choice
of sampling time which yields the strongest bound.
If we set the first derivative of $H$ with respect to $x$ equal to zero, we
find 
\begin{equation}
(1+ \epsilon)[x G' - (D-1) G] + (D-1) C_0 = 0 \, , \label{eq:minimum}
\end{equation}
which is the equation to be solved for $x_m$. We will then need to check 
that $H''(x_m) >0$ to verify that this is a minimum. 

\subsubsection{The Small $T$ Approximation}

Let us first consider quantum interest in the small $T$ limit, which is 
determined by the behavior of $G(x)$ for small $x$. Take $G$ to be an even 
function which has the approximate form
\begin{equation}
G(z) \approx C_0 - a\, z^b  \label{eq:smallx}
\end{equation}
for $0< z \ll 1$. If we substitute this form into Eq.~(\ref{eq:minimum})
and solve the resulting equation, the result is
\begin{equation}
x_m = \left[ \frac{(D-1) C_0 \,\epsilon}{a(\epsilon+1)(D-b-1)}
 \right]^\frac{1}{b} \,.
\label{eq:defxm}
\end{equation}
The bound on $\epsilon$ in this case becomes
\begin{equation}
\epsilon \geq \frac{D-b-1}{C_0} \, \left(\frac{B b}{C}\right)^{{b}/{(D-b-1)}}
 \, \left(\frac{a}{D-1}\right)^{{(D-1)}/{(D-b-1)}} \, T^{{b(D-1)}/{(D-b-1)}}
 \, . \label{eq:epsilon}
\end{equation}
As a check, we may compare this with the results of the previous subsection.
Set $D=4$ and $b=2$ in the above bound, and then set $C_0 = a = 1/\pi$, 
corresponding to the Lorentzian sampling function. Then we obtain
\begin{equation}
\epsilon \geq \frac{4}{27 \pi^2} \, \left(\frac{B}{C}\right)^2 \, T^6 \,,
\end{equation}
which is equivalent to Eq.~(\ref{eq:epsilonto0}).

We need to check that $x=x_m$ is actually a minimum of $H$. 
For our procedure to be self-consistent, we must have $b<D-1$, as may be 
seen from Eq.~(\ref{eq:defxm}). With this restriction it can be shown that 
$H''(x_m,\epsilon)>0$, for this form of $G$. There is a further constraint on $b$ 
coming from the requirement that the integrals for $C$ converge at $x=0$,
which requires that $b > \frac{1}{2}$ in two dimensions and  $b > \frac{3}{2}$ 
in four dimensions. 
In two dimensions, the lower bound on $\epsilon$ from Eq.~(\ref{eq:epsilon})
grows as $T^{{b}/{(1-b)}}$ where $\frac{1}{2} < b <1$, and hence always
grows faster than linearly for small $T$. In four dimensions, the corresponding
bound grows as $T^{{3b}/{(3-b)}}$ where $ \frac{3}{2}< b < 3$, and hence grows 
faster than $T^3$ for small $T$. The main point of this calculation 
is to show that the quantum interest effect persists even in the limit of 
arbitrarily small temporal separation of the pulses. Note that this behavior 
is also a feature of the specific example of the moving mirror.

\subsubsection{Numerical Bounds for General $T$}

\begin{figure}
\begin{center}
\leavevmode\epsfysize=7.5cm\epsffile{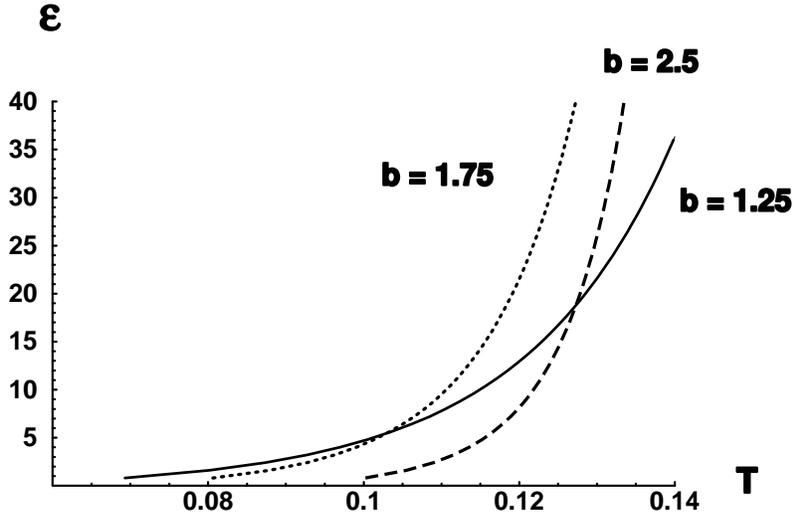}
\end{center}
\caption{
The lower bound on $\epsilon$ in two dimensions for different choices of
$b$. Here the pulse separation $T$ is in units of $|\Delta E|^{-1}$.}
\label{2DGRAPH}
\end{figure}

We can go beyond the small $T$ approximation by numerical methods. Here we
restrict our discussion to a specific choice of sampling function,
\begin{equation}
G(z) = C_0 \, e^{-|z|^b} \,,
\end{equation}
for which 
\begin{equation}
C_0 = \frac{b}{2 \Gamma(1/b)} \,.
\end{equation}
This function has the form of Eq.~(\ref{eq:smallx}) for $|z| \ll 1$, so the
bounds obtained from it will be of the form of Eq.~(\ref{eq:epsilon}) for 
small $T$. Let us first use this sampling function to study quantum interest
in two dimensions. Here Eq.~(\ref{eq:C2D}) yields
\begin{equation}
C = \frac{b^2\, \Gamma((2b-1)/b)}{24 \pi \Gamma(1/b)} \,,
\end{equation}
and Eq.~(\ref{eq:minimum}) becomes 
\begin{equation}
e^{x^b} - (1 + \epsilon)(b x^b +1)=0 \,. \label{eq:2dxm}
\end{equation}
The root of this equation is $x=x_m$. 
The upper bound on $T$ for fixed $\epsilon$, Eq.~(\ref{eq:Tbound}), may
now be expressed as 
\begin{equation}
B\, T \leq \frac{C}{b\, C_0}\, x_m^{1-b} \, (1 +b x_m^b) \,. \label{eq:Tbound2d}
\end{equation}
For a given value of $b$, we may numerically solve  Eq.~(\ref{eq:2dxm})
for fixed $\epsilon$, and check that $H''(x_m,\epsilon) > 0$. 
We then evaluate the right hand side of Eq.~(\ref{eq:Tbound2d}), 
which yields the limiting value of $T(\epsilon)$. The graph shown in 
Fig.~\ref{2DGRAPH} was obtained by inverting this relationship, and 
plotting $\epsilon(T)$. Typically,values of $b$ close to $\frac{1}{2}$ 
give the best bounds for small $T$. This
is the case that was treated analytically above. Similarly, $b \approx 1.75$
gives the best bound for larger values of $T$. Larger values of $b$ actually 
give weaker bounds for $T<T_{max}$. Some results for various values of $b$ 
are given in Figure~\ref{2DGRAPH}. Recall that Eq.~(\ref{eq:cssfT1-2d}) 
gives an upper bound on $T$ of $B\, T_{max} \approx 0.13$. 
For $T \approx T_{max}$, our sampling function
requires that $\epsilon \geq \epsilon_{min} \approx 40$.

\begin{figure}
\begin{center}
\leavevmode\epsfysize=9.5cm\epsffile{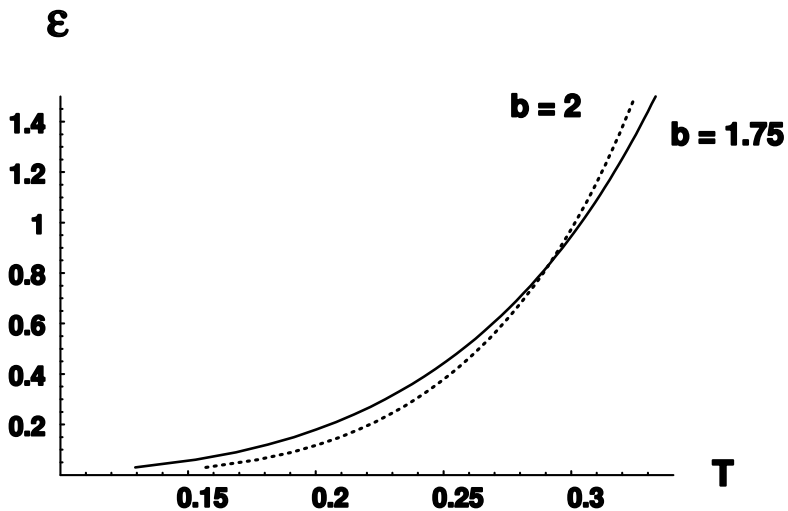}
\end{center}
\caption{
The lower bound on $\epsilon$ in four dimensions for different choices of
$b$. Here the pulse separation $T$ is in units of 
$(A/|\Delta E|)^{\frac{1}{3}}$.}
\label{4DGRAPH}
\end{figure}

A similar calculation may be performed in four dimensions. In this case
\begin{equation}
C = \frac{b^2\, (b-1)\, (2b-1)\, \Gamma((2b-3)/b)}{256 \pi^2 \Gamma(1/b)} \,,
\end{equation} 
the equation for $x_m$ is now 
\begin{equation}
3\,e^{x^b} - (1 + \epsilon)(b x^b +3) =0 \,, \label{eq:4dxm}
\end{equation}
and the bound on $T$ is
\begin{equation}
T \leq \left[ \frac{3C}{b\,B\, C_0}\, x_m^{3-b} \, (1 + \frac{1}{3}b x_m^b)
       \right]^\frac{1}{3} \,. \label{eq:Tbound4d}
\end{equation}
Again values of $b$ near the lower limit of $\frac{3}{2}$ give better bounds
for small $T$, and larger values of $b$ give better bounds for larger $T$.
Recall that Eq.~(\ref{eq:cssfT1-4d}) gives an 
upper bound on $T$ of $B^\frac{1}{3}\, T_{max} \approx 0.34$.
The bounds which arise when $b=1.75$ and $b=2$ are shown in 
Figure~\ref{4DGRAPH}. As in two dimensions, yet larger 
values of $b$ actually give weaker bounds for $T < T_{max}$.

\section{Conclusions}
\label{sec:conclusions}

We have proven that quantum states of the 
massless scalar field involving $\delta$-function 
pulses of $(-)$ and $(+)$ energy, in two and 
four-dimensional Minkowski spacetime, satisfy 
the ``quantum interest conjecture.'' This statement 
says that an energy ``loan'', i. e., $(-)$ energy, must 
always be repaid, i. e., by $(+)$ energy, with an ``interest'' 
which increases with the magnitude and/or duration of the 
``loan.'' Quantum interest is measured by the quantity $\epsilon$, 
which is the fraction by which the magnitude of the $(+)$ energy 
pulse overcompensates that of the $(-)$ energy pulse. 
A simple example of quantum interest, 
involving $\delta$-function pulses produced by a moving 
mirror, was examined in two-dimensional spacetime. There 
it was easy to see how quantum interest arose. It was the 
result of a Doppler-shifting of the subsequent $(+)$ energy 
pulse emitted by the mirror when its acceleration was 
brought to a stop in order to avoid a collision with the observer. 
The use of compactly-supported sampling functions enabled 
us to conclude that, for a fixed magnitude of the $(-)$ energy 
pulse, there must be a maximum allowed time separation 
between the pulses, in both two and four dimensions. 
This was deduced by placing the 
$(+)$ energy pulse in the region where the sampling function 
vanished. At present, we do not know what the 
optimal value of this separation might be. In two dimensions, 
we do know that it must be no smaller than 
$T = 1/(12 \, \pi |\Delta E|)$, so as to not rule out the moving 
mirror example. Another important question is which choice 
of sampling function maximizes quantum interest, for a fixed 
choice of $T$. These are unresolved problems to which we 
hope to return later. 

Moving mirrors generate more quantum interest than is 
required by our bounds. For example, $\epsilon \approx 4$ at 
$T=1/(24 \,\pi \,|\Delta E|) \approx 0.0133/|\Delta E|$, 
which is one-half of ${(T_{max})}_{mirror}$. 
This is much larger than the bounds on $\epsilon$ illustrated in 
Fig.~\ref{2DGRAPH}. It may be that the moving 
mirror state generates much more quantum 
interest than similar quantum states in which 
the pulses are produced by other mechanisms. 
Alternatively, it may be that the moving mirror state 
is close to the generic case of 
quantum interest in two dimensions. In this event, 
presumably other choices of the sampling function 
would predict a value for $\epsilon$ which is much 
closer to the moving mirror value. At present, this 
question also remains unresolved.  

Our central result is that quantum inequalities imply 
that quantum interest must exist for pulses 
in both two and four-dimensional flat spacetime. This takes 
the form of a lower bound on $\epsilon$ which is a monotonically 
increasing function of the pulse separation, for a fixed magnitude 
of the negative energy pulse.  The bound is nonzero even for 
arbitrarily small pulse separations. Once again it appears 
that nature enforces rather strict 
constraints on manipulations of $(-)$ energies. The energy 
density integrated along an inertial observer's worldline must 
be positive and the sampled energy density obeys the 
quantum inequalities. The degree of overcompensation 
by the $(+)$ energy increases with increasing pulse separation.  
In addition, there exists a maximum allowed pulse separation, 
which thereby limits the duration of the effects of the $(-)$ energy.

\vskip 0.2 in
\centerline{\bf Acknowledgements}
The authors would like to thank Michael J. Pfenning 
for useful discussions. TAR would like to thank the 
members of the Tufts Institute of Cosmology for their 
hospitality while this work was being done. This research 
was supported by NSF Grant No. Phy-9800965, 
and by a CCSU/AAUP faculty research grant.


\begin{thebibliography}{--}
\bibitem{COMMENT:CAS-EFFECT} At first sight, one might think 
that the Casimir effect is a counterexample to the quantum inequalities. 
In fact, this is not quite correct. For details, 
see Refs. \cite{FR95,PFGQI,FRWH}.
\bibitem{AREA} In this paper the area $A$ will be arbitrary. However, 
sometimes it is useful to impose the condition that all parts of the absorbing 
system come into causal contact with one another before the arrival of the 
positive energy. This implies $A < T^2$. See Ref. \cite{F91} for a discussion 
of this point.
\end{thebibliography}
\end{document}